\documentclass[twocolumn,showpacs,preprintnumbers,amsmath,amssymb,prb]{revtex4}

\usepackage{graphicx}
\usepackage{dcolumn}
\usepackage{bm}

\usepackage{amsfonts}
\usepackage{latexsym}
\usepackage{amsmath}
\usepackage{amsbsy}
\usepackage{amssymb}

\newcommand{\HoTi}{Ho$_2$Ti$_2$O$_7$ }
\newcommand{\DyTi}{Dy$_2$Ti$_2$O$_7$ }
\newcommand{\DyTins}{Dy$_2$Ti$_2$O$_7$}
\newcommand{\HoTins}{Ho$_2$Ti$_2$O$_7$}
\newcommand\degrees{\ensuremath{^\circ}}

\begin{document}


\title{Neutron Scattering Investigation of the Spin Ice State in \DyTins.}

\author{T. Fennell}\affiliation{The Royal Institution of Great Britain, 
21 Albemarle Street, London, W1S 4BS, United Kingdom}

\author{O. A. Petrenko}\altaffiliation[Now at ]{Department of Physics,
University of Warwick, Coventry, CV4 7AL, United
Kingdom}\affiliation{ISIS Facility, Rutherford-Appleton
Laboratory, Chilton, Didcot, OX11 0QX, United Kingdom}

\author{S. T. Bramwell}\email{s.t.bramwell@ucl.ac.uk}
\affiliation{Department of Chemistry, University College London,
20 Gordon Street, London, WC1H~0AJ, United Kingdom}

\author{B. F{\aa}k}\altaffiliation[Now at ]{CEA Grenoble, DRFMC/SPSMS,
38054, Grenoble, Cedex 9, France}
\affiliation{ISIS Facility, Rutherford-Appleton Laboratory,
Chilton, Didcot, OX11 0QX, United Kingdom}

\author{M. Enjalran}\altaffiliation[Now at
]{Department of Physics, Southern Connecticut State University,
New Haven, CT 06515, USA} \affiliation{Department of Physics,
University of Waterloo, Ontario, N2L 3G1, Canada}

\author{T. Yavors'kii}\affiliation{Department of Physics, University of
Waterloo, Ontario, N2L 3G1, Canada}

\author{M. J. P. Gingras}\altaffiliation[Also at ]{Canadian Institute for
Advanced Research, 180 Dundas Street West, Toronto, Ontario, M5G
1Z8, Canada} \affiliation{Department of Physics, University of
Waterloo, Ontario, N2L 3G1, Canada}

\author{R. G. Melko}\affiliation{Department of Physics, University
of California, Santa Barbara, CA 93106-9350,
USA}

\author{G. Balakrishnan}
\affiliation{Department of Physics, University of Warwick,
Coventry, CV4 7AL, United Kingdom}

\date{\today}

\begin{abstract}
\DyTi has been advanced as an ideal spin ice material.  We present a
neutron scattering investigation of a single crystal sample  of $^{162}$\DyTins.
The scattering intensity has been mapped in zero applied field in
the $h,h,l$ and $h,k,0$ planes of reciprocal space at temperatures between 0.05 K and 20
K.  The measured diffuse scattering 
has been compared with that predicted by the dipolar spin ice model.
The comparison is good, except at the Brillouin zone boundaries
where extra scattering appears in the experimental data.  It is
concluded that the dipolar spin ice model provides a successful
basis for understanding \DyTins, but that there are issues which
remain to be clarified.

\end{abstract}


\maketitle

\section{\label{intro}Introduction}
The concept of spin ice~\cite{prl1} has appeared recently as a
result of a direct analogy between the statistical mechanics of
O-H bonds in Pauling's model of water ice~\cite{pauling,giauque}
and those of the localized magnetic moments of some highly
frustrated magnets~\cite{prl1, art, science}. These materials,
typified by \HoTins~\cite{prl1}, \DyTi~\cite{art} and
Ho$_2$Sn$_2$O$_7$~\cite{kadowaki}, provide almost ideal
experimental realizations of spin ice, a magnetic ``sixteen vertex''
model~\cite{cult}. Their temperature-field phase diagrams are
exceedingly rich and have revealed several new magnetic phases and
transitions~\cite{prl1,lgcp,kagice1,kagiceprl,rod111}. The
magnetic dynamics of the spin ices also show some unique
properties including a crossover from Debye-like spin
relaxation~\cite{matsu2,shiftypeak,dirty} to quantum tunnelling dynamics~\cite{georg,snyder-condmat} at the unusually high temperature of
15 K. The current paper builds on earlier
work~\cite{prl1,newprl,kanada,kadowaki} to characterize the
neutron scattering of spin ice materials. In it, we report the
first neutron scattering images of the spin ice state of \DyTins~\cite{spinice5} which are compared with theoretical
predictions~\cite{byron,loops,Enjalran-condmat}.

In \DyTins, as well as in the other spin ice materials \HoTi and
Ho$_2$Sn$_2$O$_7$, the rare earth moments (`spins') are arranged
on a pyrochlore lattice of linked tetrahedra (see
Fig.~\ref{pyfig}). Strong crystal field
anisotropy~\cite{rosenkranz} confines  each spin as an Ising-like
doublet to the local trigonal axis that
connects the center of the elementary tetrahedron to its vertex.
With this constraint and ferromagnetic coupling, the magnetic
ground state of a single tetrahedron is defined by a ``two spins
in, two spins out'' rule, analogous to the Bernal-Fowler ice rule
that controls the proton arrangements in water ice~\cite{bernal}. This rule
minimizes  only four of
the six near neighbor magnetic interactions and so the system is
highly frustrated, sharing with ice the property of zero point
entropy~\cite{pauling,giauque,prl1,art}. In the spin ice
materials, the ferromagnetic near neighbor coupling arises from
the strong dipole-dipole interaction between neighboring
spins~\cite{byron}. If the dipolar interaction is truncated at
nearest neighbors one has the ``near neighbor'' spin ice
model~\cite{prl1,cult} (note that dipole interactions are
equivalent to exchange terms for Ising spins).  This
approximation captures the bulk properties of the spin ice
materials~\cite{prl1,art,cornelius,badoleg,fukazawa} and gives a
qualitative description of the microscopic spin correlations, as
determined by neutron scattering~\cite{prl1,newprl}.

\begin{figure}
\caption{\label{pyfig}Top: a unit cell of the pyrochlore lattice which consists
of four interpenetrating face centered cubic (fcc) Bravais sublattices.  Two spins
related by an fcc translation are shown with antiferromagnetic correlation.  
Correlations of this type give rise to the zone boundary 
scattering observed experimentally.  Bottom: the spin ice state 
on a single tetrahedron (bottom).  A 
macroscopic spin ice state is constructed by stacking such local 
arrangements on the lattice.}
\end{figure}

The subject of this paper, \DyTins, has previously been
investigated by bulk methods~\cite{art,matsu2,fukazawa,maeno}. In zero field,
the specific heat resembles a Schottky anomaly, peaking at 1.24 K
and falling to zero at lower temperatures. Unlike its \HoTi
analogue, which shows an extra peak due to freezing of the nuclear
spins~\cite{comment,newprl},  \DyTi shows no other specific heat
contributions down to 0.2 K, where the specific heat becomes
exponentially small. It has thus proved highly favorable for bulk
measurements, having been used to estimate the zero point entropy
in zero field~\cite{art}, and to demonstrate the existence of
``kagom\'e ice'', a field induced state that retains zero point
entropy~\cite{kagice1,maeno}. However, owing to its large neutron
absorption, \DyTi
has not yet been extensively studied by neutron scattering~\cite{qiu,spinice5},
the technique that was crucial in establishing the existence of
the spin ice state in \HoTins~\cite{prl1,newprl} and
Ho$_2$Sn$_2$O$_7$~\cite{kadowaki}.

To address the problem of large neutron absorption in natural dysprosium
based materials we have grown an isotopically enriched crystal of $^{162}$\DyTi
specifically for use in neutron scattering experiments. The
details of the sample and the experiments performed on it, are
given in section~\ref{nsexpsec}. The experiments were designed to
examine the nature of the presumed spin ice state that is
established when the sample is cooled in zero applied magnetic
field~\cite{art,byron}. The data were modelled using the ``dipolar'' 
spin ice model
that considers further neighbor dipole-dipole interactions as well
as a small antiferromagnetic exchange coupling that can be
estimated by fitting the susceptibility~\cite{badart} or specific
heat~\cite{byron}. 
Details of the dipolar spin ice model~\cite{byron}, as applied to \DyTins, are
given in section~\ref{mcsec} of the current paper, while the
comparison of theory and experiment is given in
section~\ref{cfsec}. The paper is concluded with a discussion of
the results, section~\ref{dissec}.

\section{Experimental}

\begin{table} [tbp]
\begin{center}
\begin{tabular}{|c|c|c|c|}
\hline
Isotope & Natural   & Sample  & $\sigma_a$ \\
        & Abundance & Content & (barn)     \\
        & (\%)      & (\%)    &            \\
 \hline \hline

 Natural     & ---       &---      & 994.(13.)\\
 \hline
 $^{156}$Dy& 0.06&$<0.01$& 33.(3.) \\
 \hline
 $^{158}$Dy &0.1&$<0.01$ & 43.(6.)\\
 \hline
 $^{160}$Dy& 2.34&0.02 &  56.(5.)\\
 \hline
 $^{161}$Dy& 19&0.47 & 600.(25.)\\
 \hline
 $^{162}$Dy& 25.5 &96.8 & 194.(10.)\\
 \hline
 $^{163}$Dy& 24.9&2.21 &  124.(7.)\\
 \hline $^{164}$Dy&28.1&0.5 & 2840.(40.)\\
 \hline Sample&---&---&207.6 \\
  \hline
\end{tabular}
\end{center}
\caption{Isotopic abundances and absorption cross sections ($\sigma_a$) of
natural dysprosium and the enriched sample used in these
experiments~\cite{munter}. } \label{dytab}
\end{table}

\subsection{Neutron Scattering}\label{nsexpsec}

Neutron scattering measurements were carried out on PRISMA at the
ISIS pulsed neutron source of the Rutherford-Appleton Laboratory.
A pulsed white neutron beam is incident on the sample. PRISMA
views a cold (95 K) methane moderator and a supermirror guide
system, providing a high flux of long wavelength neutrons
allowing measurement at relatively low $|\mathbf{Q}|$.  The
neutrons scattered from the sample are recorded in sixteen
detectors which cover a range of 16\degrees ~in scattering angle.
From the neutron time of flight, the scattering angle and the
orientation of the sample with respect to the incoming beam, the
scattered intensity can be determined as a function of the
position in reciprocal space.  By rotating the crystal about the
vertical axis, intensity maps covering a large region of the
scattering plane can be constructed.  This method is particularly
useful for studies of diffuse magnetic scattering.  The average
scattering angle $\phi$ of the sixteen detectors used for the
measurements is chosen as a compromise between flux (at a given
wavevector) and background (due to air scattering).  The
advantages of measuring the diffuse magnetic neutron scattering on
PRISMA have been clearly demonstrated using the example of
\HoTins, where the dipolar spin ice nature of the zero field spin
correlations has been unambiguously established~\cite{newprl}.  In
that and the current experiment the neutron scattering was measured
in the static approximation~\cite{collins}.

Natural dysprosium contains seven isotopes, several of which
absorb neutrons quite strongly.  This can make the detection of
weaker effects like diffuse scattering difficult or impossible. To
reduce this, an isotopically enriched sample was used. The natural
abundances, scattering lengths and cross sections for natural
dysprosium and our sample are given in Table~\ref{dytab}.  Using the $^{162}$Dy isotope the
average absorption cross-section $\bar\sigma_a$ is reduced by a factor of 4.8
compared to natural Dy. Since
absorption attenuates scattered intensity by a factor of the form
$\exp(-N\bar\sigma_a \lambda)$ (where $N$ is the number density of scatterers per 
unit volume, $\bar\sigma_a$ is the average absorption cross section, usually tabulated 
for 2200 m s$^{-1}$ neutrons, and $\lambda$ is the neutron wavelength 
scaled to the value 
used in $\bar\sigma_a$) this makes a significant difference at
long wavelengths.

A large high quality single crystal was grown using an infra-red
double mirror image furnace~\cite{geetha}.  The crystal has a
diameter of $\approx 0.4$ cm, length $\approx 1.5$ cm, and is
translucent amber-red in color. It has a cylindrical form typical
of image furnace grown crystals. The crystal was varnished into a
large copper support to ensure good thermal contact.

Using a $^3$He sorption refrigerator insert in an Orange cryostat,
the $h,h,l$ plane of reciprocal space was mapped in zero field at 20 K and $0.3$ K using
the average scattering angle $\phi=48 ^\circ$.  The same plane
was mapped at $0.3$ K and $1.3$ K using the average scattering
angle $\phi=32 ^\circ$. Using a dilution refrigerator insert in
an Oxford Instruments cryomagnet the $h,k,0$ and $h,h,l$ planes of 
reciprocal space were
mapped in zero field at $0.05$ K. These maps are shown in
Fig.~\ref{tmaps}. All parts of the sample environment in the
beam were made of aluminium (other than the copper support), which
gives rise to powder lines at $|\mathbf{Q}| = 2.70$, 3.12 and
4.40~\AA$^{-1}$.

Standard data reduction procedures were applied to transform time
of flight and angle information to reciprocal space, to correct
for absorbtion in the sample, and to normalize the data to an
absorption corrected vanadium run.  The latter corrects for the
wavelength-dependent flux profile and detector efficiency. Despite
the five-fold reduction in absorption cross section compared to natural
dysprosium, the data analysis procedure revealed that absorption
by the \DyTi crystal was still significant and needed to be taken
into account. The absorption correction was made by calculating an
attenuation coefficient ($A(s,\phi)$) by which the data could be
divided. The coefficient for a cylinder of radius $R$ was
originally derived by Sears~\cite{sears} and is
\begin{equation} \label{eq:sears}
A(s,\phi)=[1+4bs^2-0.5s^2\cos^2(\phi/2)]\times \exp(-2as),
\end{equation}
where $a=(8/3)\pi$, $b=(1-a^2)/2$, $s=\mu R$ and $\phi$ is the
scattering angle.  In this formula, $\mu$, the absorption
coefficient, depends on both the scattering cross section and the
wavelength dependent absorption cross section.

\subsection{Monte Carlo Simulations}\label{mcsec}

To model the experimental data for \DyTins, we use a
Hamiltonian appropriate for the $\langle 111 \rangle$
Ising pyrochlores:
\begin{eqnarray}
\label{eq-H}
& & H= - J \sum_{\langle (i,a),(j,b) \rangle}
{\bm S}_{i}^{a} \cdot {\bm S}_{j}^{b}  \\
& &+ D R_{\rm nn}^3 \sum_{(i,a)>(j,b)}
\left ( \frac{{\bm S}_{i}^{a} \cdot {\bm S}_{j}^{b}}{|{\bm R}_{ij}^{ab}|^3}
- \frac{3({\bm S}_{i}^{a} \cdot {\bm R}_{ij}^{ab})
({\bm S}_{j}^{b} \cdot {\bm R}_{ij}^{ab})}{|{\bm R}_{ij}^{ab}|^5} \right ),
\nonumber
\end{eqnarray}
where ${\bm S}_i^a = {\hat z}^a \sigma_i^a$ represents an Ising moment of
magnitude $|{\bm S}_i^a|=1$ at fcc lattice site $i$ and tetrahedral basis
position $a$, with its quantization axis (${\hat z}^a$) oriented along the
local trigonal axis (a member of the $\langle 111 \rangle$ set), and $\sigma_i^a = \pm 1$.
The nearest neighbor exchange energy is $J_{\rm nn} = J/3$ and
the nearest dipole-dipole strength is $D_{\rm nn}=5D/3$,
where $D=(\mu_o/4\pi)\mu^2/R_{\rm nn}^3$ and $\mu$ is the moment of the
Dy$^{3+}$ ion. ${\bm R}_{ij}^{ab}$ is the vector separation between moments ${\bm S}_i^a$ and
${\bm S}_j^b$.

The model (\ref{eq-H}), with parameters appropriate to \DyTins, orders magnetically at 0.18 K~\cite{loops}.  The real material does not show this ordering transition, but we test the hypothesis that the correlated paramagnetic state just above the transition temperature will provide an
adequate description of the spin ice state.  Equation~\ref{eq-H} was simulated  on a  pyrochlore lattice of $4\times4\times4$ cubic
cells ($N=1024$ spins) with periodic boundary conditions. The long range nature
of the
dipole-dipole interactions was properly handled by using standard Ewald
techniques for
real space dipoles~\cite{byron}. The model employed energy parameters
appropriate for \DyTins,
$J = -3.72$ K and $D = 1.41$ K~\cite{byron}.

At temperatures below 1 K or so the system is only able to access states
within the spin ice manifold.  Single spin flip dynamics require the
introduction of ice rule defects.  This means that the acceptance rate of
single spin flips becomes extremely small and ability to equilibrate the
system below 0.3 K becomes dubious~\cite{byron,loops}. In order to improve on approach to
equilibrium, 
dynamics were restored by using
a loop algorithm~\cite{loops}.  To simulate scattering data the system was
placed in the ordered groundstate found by Melko {\it et al.}~\cite{loops},
at 0.4 K.  $5\times 10 ^6$ Monte Carlo steps (MCS) per spin and $5\times 10^6$
loop attempts were made to equilibrate the system and bring it
to a paramagnetic state. Loop attempts were
therefore separated by $N=1024$ MCS. The system was then cooled to 0.3 K
where the same number of single spin flip and loop attempt steps were used for
equilibration and then data
collection (see Refs.~\onlinecite{byron,newprl,melko1} for more
details).

There are 12 symmetry related ordered stuctures and to maintain ergodicity each 
was used as a start point for a simulation.  $I({\bm Q})$ was calcuated at $T=0.3$K from
200 independent spin configurations derived from each starting structure
using the formula~\cite{newprl}
\begin{equation}
I({\bm Q}) \propto \frac{|f(Q)|^2}{N} \sum_{(i,a),(j,b)}
{\rm e}^{\imath{\bm Q}\cdot{\bm R}_{ij}^{ab}}
\langle {\bm S}_{i,\perp}^a \cdot {\bm S}_{j,\perp}^b \rangle \ ,
\end{equation}
where $\langle ... \rangle$ denotes a thermal average, ${\bm S}_{i,\perp}^a$
is the spin component perpendicular to the scattering wave vector ${\bm
Q}$ and
$f(Q)$ is the magnetic form factor for Dy$^{3+}$. The
Monte Carlo results for $I({\bm Q})$ are compared to the experimental data
in the $h,h,l$ and $h,k,0$ planes below. The simulations are nominally
at 0.3 K,
whereas the experimental data was collected at either 0.3 K ($h,h,l$
plane) or 0.05 K ($h,k,0$ plane).  We consider it valid to make
these comparisons because at temperatures below 0.4 K the
susceptibility~\cite{fukazawa}, neutron scattering intensity, $I({\bm Q})$
and specific heat~\cite{art} do not change significantly with temperature: the system
essentially appears to be frozen with negligible thermal spin fluctuations.

\section{Results}\label{cfsec}

In this section, the observed magnetic diffuse scattering is
presented and compared to Monte Carlo simulations where possible.
The temperature dependence of the scattering in the $h,h,l$ plane is
discussed first (section~\ref{resecT}), in order to establish the
temperature regimes of interest.  In the following sections
(\ref{resechk0} and \ref{resechhl}) the scattering intensity
measured at base temperature is compared to Monte Carlo
simulations. Finally, section~\ref{reseczones} is concerned with
features in the experimental scattering which are not reproduced
by the model.

\subsection{Temperature Dependence of the Diffuse
Scattering}\label{resecT}

The temperature dependence of the diffuse scattering due to the
development of ice rule fulfilling spin correlations was sampled
at four temperatures for the $h,h,l$ plane (20, 1.3, 0.3 and 0.05 K,
see Fig.~\ref{tmaps}). An important point that can be obtained
from these measurements is that the diffuse scattering does evolve
with temperature. As the sample is cooled, no diffuse scattering
characteristic of spin ice correlations has built up by 20 K. This
is entirely expected as the bulk susceptibility shows that \DyTi
is paramagnetic at this temperature~\cite{fukazawa}. At $1.3$ K
strong diffuse scattering has appeared. The form is closely
similar to that of the near-neighbor spin ice as simulated in
Ref~\onlinecite{newprl}. Between $1.3$ and $0.3$ K the spin correlations
evolve and additional features appear in the scattering pattern,
as discussed below.  The scattering at
0.05 K throughout the $h,h,l$ plane (not illustrated here) is not
significantly different to that at 0.3 K.

\begin{figure*}
\caption{\label{tmaps}(Color online) \DyTins: Diffuse scattering in the $h,h,l$ plane 
measured at $20$ K (top left), $1.3$ K (top right) and $0.3$ K 
(bottom left, with zone boundaries illustrated).  The sharp, intense spots are nuclear Bragg reflections 
with no magnetic intensity.  The map at 20 K was recorded using an average 
scattering angle of $\phi = 48$\degrees~ which gives stronger arc-shaped 
artefacts for small
$|\mathbf{Q}|$ values compared to $\phi = 32$\degrees.  The result of the 
Monte Carlo simulation of the
dipolar spin ice model at 0.3 K is shown at bottom right.
}
\end{figure*}

\subsection{Scattering in the $h,k,0$ plane}\label{resechk0}

For the $h,k,0$ plane, data has only been collected at 0.05 K.  The simulation 
and the experiment seem to agree
very well indeed. The two scattering maps are shown in
Fig.~\ref{hk0maps}. A comparison by a cut through the data is
shown in Fig.~\ref{slices}. The Monte Carlo pattern has been fitted
to the experimental diffuse scattering using a flat background and
an overall intensity scale parameter.

\begin{figure*}
\caption{\label{hk0maps}(Color online) \DyTi at 0.05 K: experimental scattering
intensity in the $h,k,0$ plane (left), compared with that calculated
from Monte Carlo simulations (right).}
\end{figure*}

\subsection{Scattering in the $h,h,l$ plane}\label{resechhl}

For the $h,h,l$ plane direct comparison of the experimental and
simulated scattering maps reveals both close similarities and
significant differences.  The two scattering patterns are shown 
in Fig.~\ref{tmaps}.   It is apparent that whilst the experimental
pattern does contain the features of the dipolar spin ice
scattering (for example the large diffuse feature at $0,0,3$) it
also contains features that are are not reproduced by the
simulation. These extra features are concentrated around the
boundaries of the Brillouin zones (the position of the zone boundaries are illustrated in Fig.~\ref{tmaps}).


The similarities and differences of the observed and calculated
patterns are more clearly emphasized in Fig.~\ref{slices}.  Here
slices through the data have been made.    The agreement along $h,h,0$ and $0,0,l$ is good. The slice along $h,h,2.3$ is chosen
to highlight the differences and it can be seen that where the
slice crosses the zone boundary the form of the calculated and
experimental intensities is significantly different.

\begin{figure*}
\caption{\label{slices}\DyTi at 0.3 K: comparison of experimental
and Monte Carlo data for $h,k,0$ and $h,h,l$ planes.  The comparison
in the $h,k,0$ plane is shown at top left using a slice along
$3,k,0$.  The Monte Carlo data was compared to the experimental
data by adjusting an overall intensity scale parameter and single 
background parameter
in order to give the best fit. The comparisons for the $h,h,l$ plane
are slices along $h,h,0$ (top right), $0,0,l$ (bottom left) and
along $h,h,2.3$ (bottom right). For slices 2 and 3 the Monte Carlo
data were compared using an overall intensity scale parameter
 and sloping background,
slice 4 was compared with an overall intensity scale parameter
 and constant background.
The slice positions in the scattering plane are indicated in the
insets, and for the $h,h,l$ plane by numbers 2-4. Shaded areas in
the slices are Bragg peaks from the sample and were excluded from
the fitting.}
\end{figure*}

\subsection{Zone Boundary Features}\label{reseczones}

One possible explanation for the zone boundary features is that
they are due to phonons. However, the following facts suggest that
the intensity is magnetic: (1) the scattering appears as
temperature is decreased; the converse would be expected for a
phonon; (2) the experimentally observed features fall off with
increasing $|\mathbf{Q}|$, which is not expected for phonons.

The second point can be demonstrated by comparison of the
intensity of the zone boundary features with the magnetic form factor of
Dy$^{3+}$. If the intensity at equivalent points on successive
zone boundaries is compared it would be expected to follow the
form factor if the scattering process is magnetic.  This is found to 
be the case.  On the basis of 
this comparison it seems entirely reasonable to attribute the zone boundary
features to a magnetic correlation.

\section{Discussion}\label{dissec}

The dipolar spin ice model apparently provides a very successful
model for the zero field spin correlations in Dy$_2$Ti$_2$O$_7$.
However, some additional spin correlations do occur which are not
captured by the model. The model is very successful in describing
the bulk and microscopic data for Ho$_2$Ti$_2$O$_7$~\cite{newprl},
and the bulk data for Dy$_2$Ti$_2$O$_7$~\cite{byron}.  One might
therefore ask, how this further
restriction of the dipolar spin ice manifold is not manifested in
the fit to the heat capacity?

In answer to this question, it should be noted that the fit to the
heat capacity~\cite{byron} is not perfect: the experiment and
simulation differ in detail. The fact that the zone boundary
features are very diffuse in reciprocal space and run all around
the zone boundaries, suggests that the associated spin
correlations are short ranged and isotropic.  A small difference
between the experimental and theoretical heat capacities is
therefore understandable since the extra spin correlations need
not make a major contribution to the entropy. We therefore
conclude that the neutron scattering and specific heat
data~\cite{art,byron} are consistent.

A clue to the origin of the extra spin correlations is the fact
that they are much less evident in \HoTins~\cite{newprl,kanada},
even if they cannot be excluded completely within experimental
error. According to the model, the principal difference between
\HoTi and \DyTi is the more significant antiferromagnetic
exchange term in the spin Hamiltonian of \DyTi ($J=-3.72$ K
 for \DyTi compared to $J=-1.56$ K for \HoTins).  This exchange 
term was invoked to fit bulk measurements~\cite{badart,fukazawa,byron}, 
which are
not sensitive to the details of a Hamiltonian in the way that
diffuse neutron scattering is. Hence the extra scattering might
reveal that extra terms are required in the model
Hamiltonian: for example, it might be necessary to include further
neighbor exchange interactions.  Further theoretical work is required to 
distinguish various possibilities.

Zone boundary scattering has 
recently also been observed in the $h,k,0$ plane for ZnCr$_2$O$_4$~\cite{director} and the $h,h,l$ plane for ZnFe$_2$O$_4$~\cite{kamazawa} (both
frustrated antiferromagnets with the spinel structure).
We suggest that, while the three magnets are quite different, the zone boundary
features may also be a generic feature of frustrated magnets due to the 
formation of local spin clusters, for example those developing in ZnCr$_2$O$_4$~\cite{director,protectorate-exp} which might be at the origin of the ``quantum protectorate'' state invoked in that system.  As noted by Kamazawa~{\it et
al.}~\cite{kamazawa}, zone boundary scattering implies correlations between the spins related
by fcc translations (as illustrated in Fig.~\ref{pyfig}).  These spins are third neighbors and interactions at
this distance will clearly be important in stabilizing a hexagonal cluster.
Simulations suggest that other clusters can also be stabilized
depending on the details of the Hamiltonian~\cite{tomntarasunp}.    


In conclusion, we have presented a systematic investigation of the
diffuse scattering of \DyTi in zero magnetic field.  We have
observed three regimes of behavior.  At 20 K the spins are
paramagnetic, without spin ice correlations,
which should help interpret recent bulk
susceptibility measurements of the dynamical
properties~\cite{matsu2,fukazawa,shiftypeak}, which have been
discussed in terms of both correlated and uncorrelated dynamical
processes~\cite{shiftypeak,georg,snyder-condmat}. At 1.3 K the
spin correlations are similar to those of the near neighbor spin
ice model, suggesting that further neighbor dipole interactions
are not strongly relevant in this temperature range. At 0.3 K the
scattering is modified to correspond quite closely to the frozen
state of the dipolar spin ice model, with effective infinite range
interactions.  However, in this temperature range we have observed
additional scattering features around the zone boundary which for
the time being remain unexplained.  Further calculations are in
progress to clarify their origin.

\section{Acknowledgements}

We thank the sample environment team at ISIS, in particular
Richard Down and Dave Bates, and Martyn Bull for assistance on
PRISMA. TF acknowledges the EPSRC for funding, and Andrew Harrison
for hospitality in Edinburgh whilst this work was written up. Work
done at the University of Warwick was funded by an EPSRC grant.
Work done at the University of Waterloo has been supported by the
NSERC of Canada, the Canada Research Chair Program, the Province
of Ontario, and Research Corporation.

\bibliography{dypaper2}

\begin{thebibliography}{41}
\bibitem[{pro()}]{protectorate-exp}
\bibinfo{note}{In ZnCr$_2$O$_4$~\cite{director} hexagonal loops of spins
  organize into weakly interacting antiferromagnetic clusters. The fluctuation
  of the directors of the hexagons governs the low temperature dynamics. This
  state is termed a ``protectorate'' as it appears to be a stable state of
  matter in a strongly correlated system~\cite{laughlin}.}
\end{thebibliography}

\end{document}